\documentclass[aps,pra,floatfix,onecolumn]{revtex4}

\usepackage{amsmath}
\usepackage{amssymb}
\usepackage{graphicx}

\begin{document}

\title{Efficient quantum computing using coherent photon conversion}

\author{N.~K.~Langford$^{1,2,3,*}$, S.~Ramelow$^{1,2}$, R.~Prevedel$^{1,4}$, W.~J.~Munro$^5$, G.~J.~Milburn$^{6,1}$ \& A.~Zeilinger$^{1,2,*}$}

\affiliation{$^1$ Vienna Center for Quantum Science and Technology (VCQ), Faculty of Physics, University of Vienna, Boltzmanngasse 5, A-1090, Vienna, Austria \\
$^2$ Institute for Quantum Optics and Quantum Information (IQOQI), Austrian Academy of Sciences, Boltzmanngasse 3, A-1090, Vienna, Austria \\
$^3$ Clarendon Laboratory, Department of Physics, University of Oxford, Parks Road, Oxford, OX1 3PU, UK \\
$^4$ Institute for Quantum Computing, University of Waterloo, Waterloo, N2L 3G1, ON, Canada \\
$^5$ NTT Basic Research Laboratories, NTT Corporation, 3-1 Morinosato-Wakamiya, Atsugi, Kanagawa 243-0198, Japan \\
$^6$ Centre for Engineered Quantum Systems, University of Queensland, St Lucia, 4072, Queensland, Australia
}

\begin{abstract}
Single photons provide excellent quantum information carriers, but current schemes for preparing, processing and measuring them are inefficient.  For example, down-conversion provides heralded, but randomly timed single photons, while linear-optics gates are inherently probabilistic.  Here, we introduce a deterministic scheme for photonic quantum information.  Our single, versatile process---\emph{coherent photon conversion}---provides a full suite of photonic quantum processing tools, from creating high-quality heralded single- and multiphoton states free of higher-order imperfections to implementing deterministic multiqubit entanglement gates and high-efficiency detection.  It fulfils all requirements for a scalable photonic quantum computing architecture.  Using photonic crystal fibres, we experimentally demonstrate a four-colour nonlinear process usable for coherent photon conversion and show that current technology provides a feasible path towards deterministic operation.  Our scheme, based on interacting bosonic fields, is not restricted to optical systems, but could also be implemented in optomechanical, electromechanical and superconducting systems which exhibit extremely strong intrinsic nonlinearities.
\end{abstract}

\maketitle

Photonic qubits were used in the earliest demonstrations of entanglement~\cite{ClauserJF1978a} and also to produce the highest-quality entanglement reported to date~\cite{BarreiroJT2005a,FedrizziA2007a}.  One of the key challenges for photonic quantum information processing (QIP) is to induce strong interactions between individual photons, which cannot be realised with standard linear optical components.  The scheme proposed by Knill, Laflamme and Milburn for linear optics quantum computing~\cite{KnillE2001a,KokP2007a} (LOQC) managed to sidestep this problem by using the inherent nonlinearity of photodetection and nonclassical interference to induce effective nonlinear photon interactions nondeterministically.  Alternatively, in the one-way picture of quantum computing, the required nonlinearities are replaced by off-line probabilistic preparation of special entangled states followed by detection and feed-forward~\cite{RaussendorfR2001a,WaltherP2005a,PrevedelR2007a}.

Nonlinear optics quantum computing (NLOQC) takes a different approach by directly using intrinsic nonlinearities to implement multiphoton interactions.  NLOQC schemes have been proposed using different types of optical nonlinearities, including cross-Kerr coupling~\cite{MunroWJ2005a,NemotoK2004a} and two-photon absorption~\cite{PittmanTB2004c}.  Since then, more complete multimode analyses of the cross-Kerr NLOQC schemes suggest that they cannot in fact produce phase shifts large enough for NLOQC because of spectral correlations created between the interacting fields~\cite{ShapiroJH2006a,ShapiroJH2007a,Gea-BanaclocheJ2010a}.  Other work, however, shows that these difficulties can be circumvented in the related case of strong $\chi^{(2)}$ interactions by carefully engineering the phase-matching conditions~\cite{LeungPM2009a}.

Here, we introduce an alternative nonlinear process---\emph{coherent photon conversion} (CPC)---based on creating an enhanced, tunable nonlinearity by pumping higher-order nonlinear interactions with bright classical fields, allowing the potential for multiparty control of coherent nonlinear dynamics.  A simple example of CPC gives an effective quadratic (three-wave mixing) nonlinearity by pumping one mode of a four-wave mixing interaction.  We show that this example is an extremely versatile process which provides a range of useful photonic QIP tools, including deterministic two-qubit entangling gates based on a novel form of effective photon-photon interaction, high-quality heralded sources of multiphoton states and robust and efficient single-photon detection.  Such tools are valuable building blocks in many optical quantum-enabled technologies.  In particular, we outline a new approach for optical quantum information processing based on CPC that fulfills all of the DiVincenzo criteria for a viable implementation of quantum computation~\cite{DiVincenzoDP1998a}.  Moreover, it can also be used to produce heralded multiphoton entanglement, create optically switchable quantum circuits, convert arbitrary input states into high-purity, probabilistic single- and multiphoton Fock states, and implement an improved form of down-conversion with higher emission probabilities and lower higher-order terms.

In standard photonic systems, cubic (four-wave mixing) nonlinearities in optical fibres have produced some of the highest-brightness photon pair sources~\cite{FulconisJ2005a,AlibartO2006a}, and the available precise dispersion engineering and fibre structuring technologies have allowed optimisation of these sources to produce ultrabright high-purity heralded single photons~\cite{HalderM2009a,CohenO2009a}.  Here, we carry out proof-of-principle experiments using $\chi^{(3)}$ photonic crystal fibres which demonstrate the four-mode interaction underlying CPC via photon doubling of a weak input state.  Critically, we show that we can tune and enhance the effective $\chi^{(2)}$ interaction strength by varying the classical pump power.  We also use these experiments to characterise the strength of the corresponding CPC interaction and outline how to reach the deterministic regime with current technology.

\section{Coherent Photon Conversion}

The fundamental process underlying coherent photon conversion is a nonlinear interaction between $m$ bosonic modes which coherently converts single excitations in some of the modes (depending on the precise form of the interaction), into single excitations in the remaining modes.  A key principle of CPC is that this basic nonlinearity can in turn be generated by pumping some modes of a higher-order nonlinearity with strong classical fields.  This induces an effective coupling between the quantum modes which can be tuned (and enhanced) by the classical pumps.  Indeed, it may be possible to produce an effective interaction which is stronger than naturally occurring couplings of the same form.  A similar effect is achieved in photon pair sources based on four-wave mixing in photonic crystal fibres, where very high pair rates are achieved with very low pump powers~\cite{FulconisJ2005a,AlibartO2006a}.  To illustrate the potential of CPC, here we focus on a novel case which has very interesting properties for applications in quantum optics.

Consider a standard four-wave mixing interaction involving four distinct frequency modes:
\begin{equation}
H = \gamma a b^\dag c^\dag d + \gamma^\ast a^\dag bc d^\dag,
\end{equation}
where the strength, $\gamma$, arises from the third-order ($\chi^{(3)}$) nonlinearity.  Pumping one of the modes ($d$) with a bright classical beam, $E$, yields the effective second-order interaction:
\begin{equation}
\tilde{H} = \tilde{\gamma} a b^\dag c^\dag + \tilde{\gamma}^\ast a^\dag bc,
\end{equation}
where $\tilde{\gamma}\propto\gamma E$.  This now looks like a standard three-wave mixing Hamiltonian.  To explain the basic CPC operation, here we use a simple single-mode ``time-evolution'' model where the modes satisfy energy conservation, $\Delta \omega = \omega_a - \omega_b - \omega_c + \omega_d = 0$, and we for the moment ignore any higher level of sophistication introduced by imperfect phase-matching and photon loss.  

The key to understanding how this process works, and its potential, is to look at its effect on input \emph{Fock} states of the form $|n_a n_b n_c\rangle$.  In particular, the Hilbert space defined by the set of states $\tilde{H}^j |n_a n_b n_c\rangle$ (for all integers $j$) is a greatly restricted subspace of dimension $(n_a{+}\min(n_b,n_c){+}1)$.  Consequently, if the quantum state starts in a product number state, it will evolve entirely within this finite subspace and will therefore exhibit the collapses and revivals in individual population elements which are characteristic of coherent quantum processes.  The most important example of this for our scheme is the case which evolves within a novel two-dimensional subspace, $\{ |100\rangle , |011\rangle \}$: specifically, $\tilde{H}|100\rangle \propto|011\rangle$ and $\tilde{H}^2|100\rangle \propto|100\rangle$. Thus, the coupling induced by the Hamiltonian then drives Rabi-like oscillations between the two basis states, i.e., given $|100\rangle$ as an input, the output state evolves as:
\begin{equation}
|\psi(t)\rangle = \cos \left( \Gamma t \right) |100\rangle + i \frac{\tilde{\gamma}}{\left| \tilde{\gamma} \right|} \sin \left( \Gamma t \right) |011\rangle,
\label{eq:evolution-onephoton}
\end{equation}
where $\Gamma = \left| \tilde{\gamma} \right| / \hbar$.

Interestingly, standard single photon up-conversion, a special case of CPC (indeed the simplest case), is one of the small subset of CPC processes with purely classical analogues: if a classical input field is used, then, provided this ``pump'' field remains undepleted throughout, the input field will undergo complete coherent oscillations between the two frequency modes.  By contrast, if a classical input is used in the above example, then no coherent oscillations will be observed---the output is the well-known two-mode squeezed state of a parametric down-conversion source.  In other words, in most cases a key element of CPC operation is the use of quantised inputs.

\section{Tools for Quantum Computing}

\begin{figure}
\begin{center}
\includegraphics[width=0.55\columnwidth]{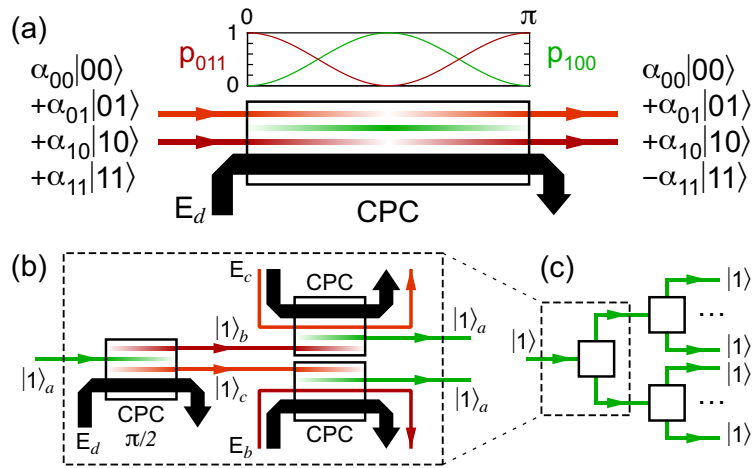}
\end{center}
\caption{Fulfilling the DiVincenzo criteria with CPC. (a) Deterministic controlled-phase gate. A ``$\pi$'' CPC interaction ($t{=}\pi/\Gamma$) is a novel type of effective photon-photon interaction, which implements an entangling CZ gate between two logical states of frequency non-degenerate photons (e.g., polarisation or spatial encoding). (b) Scalable element for deterministic photon doubling. A ``$\pi/2$'' CPC interaction ($t{=}\pi/2\Gamma$) can be used both to convert any single-photon source into a good source of multiphoton states and to perform high-efficiency, low-noise detection at any wavelength. (c) Deterministic photon doubling cascade.  The scalable photon doubler from (b) (depicted by the symbol shown in the inset) can be directly chained to create a deterministic cascade for either multiphoton state preparation or detection enhancement.}
\label{fig:divincenzo-criteria}
\end{figure}

\begin{figure}
\begin{center}
\includegraphics[width=0.6\columnwidth]{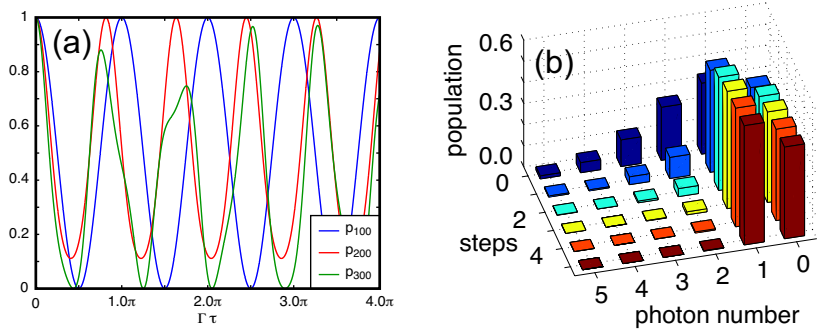}
\end{center}
\caption{Heralded single-photon source. (a) Evolution of $|n_a00\rangle$ populations under the CPC interaction for $n_a{=}1$, $2$ $\&$ $3$. (b) Number-state populations after each filtering step for $t{=}\pi/\Gamma$, giving $|1\rangle_a$. Combined with a single photon-doubling step and given a weak coherent state with $|\alpha|^2=1.5$ as input, this scheme gives a heralded single-photon source with production efficiency of ${\sim}56\%$ and virtually no higher-order photon-number terms (${<}0.3\%$) in only $5$ steps.}
\label{fig:heralded-single-photon}
\end{figure}

The DiVincenzo criteria describe the basic conditions for a viable implementation of quantum computing~\cite{DiVincenzoDP1998a}.  The major unresolved challenges for photonic quantum information are good multiphoton sources, reliable multiqubit interactions, and robust, high-efficiency single-photon detection.  We show here that CPC provides tools to solve all three of these issues (Figs~\ref{fig:divincenzo-criteria} and \ref{fig:heralded-single-photon}), all derived from a single process just by choosing different interaction strengths.

Figure~\ref{fig:divincenzo-criteria}a illustrates how CPC directly implements a two-qubit controlled-Z (CZ) gate between the photons in the two modes $b$ and $c$.  The key insight is that CPC, like any coherent process which cycles between two orthogonal states, exhibits geometric (Berry's) phase effects~\cite{BerryMV1984a,PancharatnamS1956a, RauchH1975a} (cf.\ the all-optical switch demonstrated in Ref.~\cite{VanDevenderAP2007b}).  Therefore, for $t = \pi/\Gamma$, an input state $|011\rangle$ will undergo a full oscillation and pick up a $\pi$ phase shift, giving the final state $-|011\rangle$.  Because this phase shift only occurs when \emph{two} single photons are present, this \emph{controlled} phase shift can be used to implement a maximally entangling CZ gate with 100\% efficiency.  Note that this is a truly non-classical geometric phase which has no equivalent with classical input states.  This CZ gate can also be switched very fast optically (by switching the bright classical pump beam in and out, cf.\ Ref.~\cite{VanDevenderAP2007b}), allowing the fast, real-time ``rewiring'' of optical quantum circuits.  This may have application in various adaptive quantum schemes such as quantum phase estimation or adaptive quantum algorithms and might be particularly useful in wave-guide and integrated-optics architectures~\cite{PolitiA2008a}.

If the input state is only allowed to undergo half an oscillation ($t = \pi/2\Gamma$), a single photon can be converted \emph{coherently and deterministically} (i.e.\ with 100\% efficiency) into two single photons---a deterministic photon doubler~\cite{KoshinoK2009a} (or alternatively a deterministic two-photon absorber in the reverse direction).  Figure~\ref{fig:divincenzo-criteria}b illustrates one method for implementing a scaleable photon doubler, allowing them to be chained together to create an arbitrary number of photons (see Supplementary Information).  This efficient photon doubling cascade (Fig.~\ref{fig:divincenzo-criteria}c) can be used to create a high-quality, scaleable source of multiphoton states from any source of genuine single photons (on-demand or heralded).  Note that the photon doubler can also be used in conjunction with existing methods to create arbitrary, heralded (perhaps also nonlocally prepared) entangled Bell-type two-photon~\cite{BellJS1964a} and GHZ-type three-photon~\cite{GreenbergerDM1990a} states (see Supplementary Information). Indeed, these tools can directly implement the encoding step for a simple 9-qubit error-correction scheme~\cite{ShorPW1995a}.

The same photon doubling cascade (Fig.~\ref{fig:divincenzo-criteria}b) can also be used to perform high-efficiency, low-noise detection with real-world noisy, inefficient detectors (the photon doubler implements the quantum copier from~\cite{DeuarP1999a}).  The cascade creates $n$ copies of any single photon arriving at the detector and a detection event can then be defined to be a $k$-fold coincidence between any $k$ detectors at the output.  By choosing $n$ and $k$ appropriately, we can simultaneously boost the detector efficiency and reduce the noise from dark counts (see Supplementary Information).  Interestingly, this technique can also produce marked improvements in detector characteristics, even when the photon doubling operates at lower efficiencies ($\eta_{\rm dbl} {<} 1$).  Indeed, photon-doubling can be used to improve overall detector efficiencies if $\eta_{\rm dbl} > 1/(2{-}\eta)$, where $\eta$ is the single-detector efficiency.  For example, for $\eta= 50\%$, the doubling efficiency would need to be greater than 2/3.  Moreover, if the remaining mode can be detected as well, the final efficiency can be improved for any value of $\eta_{\rm dbl}$.

Finally, CPC can also be used to create a high-fidelity source of heralded single photons which could be used to seed the efficient photon doubling cascade described above.  As noted previously, higher-order input states of the form $|n_a00\rangle$ will evolve within a restricted, $(n_a{+}1)$-dimensional Hilbert space.  As with the qubit case, this leads to coherent oscillations of population, as illustrated in Fig.~\ref{fig:heralded-single-photon}a for $n_a=1$, 2 and 3, but their complexity increases rapidly with larger $n_a$, because the evolution is governed by an increasingly complicated distribution of eigenfrequencies (e.g., $n_a=3$ in Fig.~\ref{fig:heralded-single-photon}a; see Supplementary Information for details).  As more competing frequencies come into play, for higher orders these oscillations are characterised by collapses and revivals in the input state population at often irregular times.  Remarkably, these frequencies are incommensurate with the frequencies from other orders, so the revivals occur at different times for different input states (Fig.~\ref{fig:heralded-single-photon}a).

Consider therefore an input state in mode $a$ which is a superposition (or mixture) of different $n_a$, e.g., $|\psi(0)\rangle_a {=} \sum_j \alpha_j |j\rangle_a$ or $|\psi(0)\rangle_a {=} |\alpha\rangle_a$ (a ``classical'' coherent state).  When the $|100\rangle$ term has undergone one complete oscillation (i.e., same as the CZ gate, $t{=}\pi/\Gamma$), all other terms will, with non-zero probability, have converted into states with photons in modes $b$ and $c$, which can be rejected via spectral filtering.  Applying this process repeatedly will suppress all contributions from other orders, leaving only the $|1\rangle_a$ state with a finite probability (Fig.~\ref{fig:heralded-single-photon}b).  (Note: by detecting the dump port of the filtering step with high efficiency and rejecting events which lead to clicks in these arms, this acts like a pure Fock-state filter~\cite{ZeilingerA1992a}.)  By combining this with a single coherent photon-doubling step, it then becomes a \emph{heralded} single-photon source.

Currently, spontaneous parametric down-conversion (SPDC) and spontaneous four-wave mixing (SFWM) provide the best available sources of heralded single photons, but the performance and achievable rates of these sources are intrinsically limited by the effects of higher-order photon-number terms~\cite{BarbieriM2009a}.  By contrast, given a simple weak coherent state with $|\alpha|^2=1.5$ as input, for example, our scheme provides heralded single photons with production efficiencies of ${\sim}56\%$ and virtually no higher-order photon-number terms (${<}0.3\%$) in only 5 steps (Fig.~\ref{fig:heralded-single-photon}b).

Using similar principles, CPC can also be used to probabilistically create other small Fock states with high fidelity (e.g., $|200\rangle$) and to implement an improved form of down-conversion which can provide substantially higher pair-emission probabilities with much higher ``heralded'' state fidelity than a standard down-conversion source with comparable emission rates (see Supplementary Information).

\section{Experimental coherent photon conversion}

\begin{figure}
\begin{center}
\includegraphics[width=0.8\columnwidth]{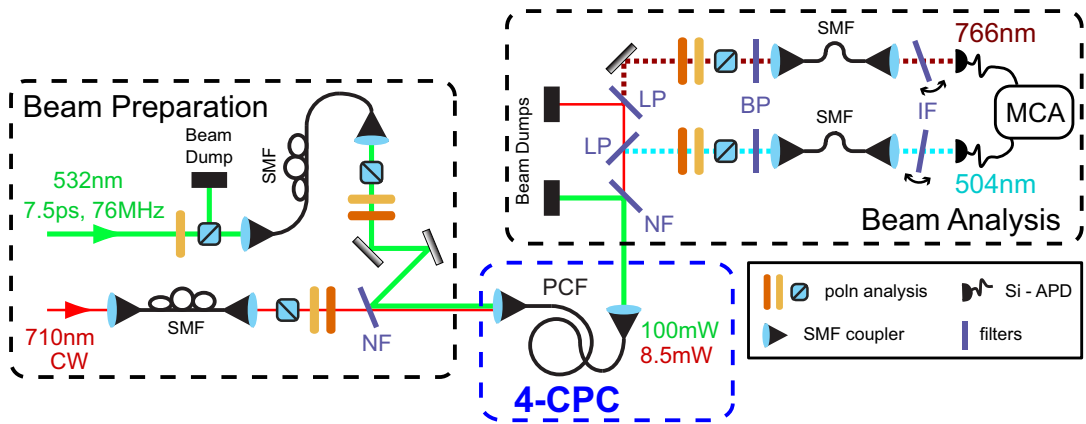}
\end{center}
\caption{Schematic of the experiment. The nonlinear medium is a standard commercial, polarisation-maintaining photonic crystal fibre (PCF: 1.8 $\mu$m core, nonlinearity ${\sim} 95 \,({\rm W\,km})^{-1}$). The $\chi^{(3)}$ nonlinearity is pumped by a frequency-doubled pulsed neodymium vanadate laser ($\text{Nd:YVO}_4$: 532 nm, 7.5 ps, 76 MHz) to create the desired tunable effective $\chi^{(2)}$ nonlinearity.  A continuous-wave (cw), external-cavity diode laser (710 nm, ${\sim} 2 {\times} 10^5$ photons per $\text{Nd:YVO}_4$ pulse) provides the input state in mode $a$ which we use to characterise the strength of the CPC interaction. From estimated dispersion for the PCF, birefringent phase matching is satisfied for the following four-mode interaction: 532 nm (H) + 710 nm (H) $\rightarrow$ 504 nm (V) + 766 nm (V), where H and V denote horizontal and vertical polarisation, respectively.  The 532 nm and 710 nm input beams are spatially filtered with single-mode fibres (SMF) before being combined on a notch filter (NF) and coupled into the PCF. The beams emerging from the output are then spectrally separated using a range of filters (NF, LP: long-pass, BP: band-pass) and a simple monochromator (a rotating interference filter, IF), passed through polarisation analysers, and finally analysed in coincidence using time-to-amplitude conversion and a multichannel analyser (MCA).}
\label{fig:experiment}
\end{figure}

There are many different media that can be used to provide the type of $\chi^{(3)}$ nonlinear interaction required for CPC.  Some examples which are very promising for strong $\chi^{(3)}$ interactions are standard optical fibres~\cite{Agrawal}, photonic crystal fibres~\cite{KnightJC2003a}, silicon waveguides~\cite{FosterMA2006a} and EIT media~\cite{HamBS1997a,BrajeDA2004a}.
Although the $\chi^{(3)}$ nonlinearity for a given material is normally much weaker than the $\chi^{(2)}$ nonlinearity for the same material (often by many orders of magnitude), the enhancement by the classical field can, for a sufficiently strong pump, result in an \emph{effective} $\chi^{(2)}$ nonlinear interaction which is stronger than the available natural $\chi^{(2)}$ interaction.  There are several key advantages to using such a pumped $\chi^{(3)}$ interaction to produce an effective lower-order nonlinearity.  Firstly, materials with inversion symmetry have no $\chi^{(2)}$ nonlinearity ($\chi^{(2)}=0$), but all materials possess some $\chi^{(3)}$ nonlinearity.  Using the classical pump creates a quadratic nonlinearity which is tunable and no longer limited by fixed material properties.  Finally, and perhaps most importantly, conservation of energy allows the four-mode interaction to take place between nearly degenerate frequency modes, which makes CPC compatible with standard telecom-band fibre-based implementations, unlike standard $\chi^{(2)}$ interactions, where the pump must by definition be roughly double the frequency of the other two photons.

There are several possible types of four-wave mixing interactions, all of which occur simultaneously according to various selection rules given by symmetries in the $\chi^{(3)}$ nonlinear susceptibility tensor and governed by phase-matching conditions.  Some of these processes, like cross-Kerr and self-Kerr phase modulation, are automatically phase-matched, but they do not inhibit coherent photon conversion, because they only modify the phase-matching conditions of the target process, rather than competing with it.  Of the other processes, careful engineering of the scheme and the nonlinear medium should be able to ensure that only one is phase-matched and the others can be neglected.

In these first proof-of-principle experiments, we aim to demonstrate the viability of the nondegenerate four-mode nonlinear interaction underlying the above CPC process. In particular, we wish to demonstrate the tunability
of the effective $\chi^{(2)}$ interaction and to estimate its efficiency to determine whether current technology offers a feasible path towards the deterministic regime.  To provide the required $\chi^{(3)}$ nonlinearity, we use a standard commercial, polarisation-maintaining nonlinear photonic crystal fibre (PCF) and pump it with a 532 nm pulsed laser.
We use weak coherent states from a 710 nm diode laser in mode $a$ and characterise the CPC interaction strength via the resulting double-pumped down-conversion pair source.  With horizontally polarised input photons at 532 nm and 710 nm, birefringent phase matching leads to vertically polarised output photons at 504 nm and 766 nm.

\begin{figure}
\begin{center}
\includegraphics[width=\columnwidth]{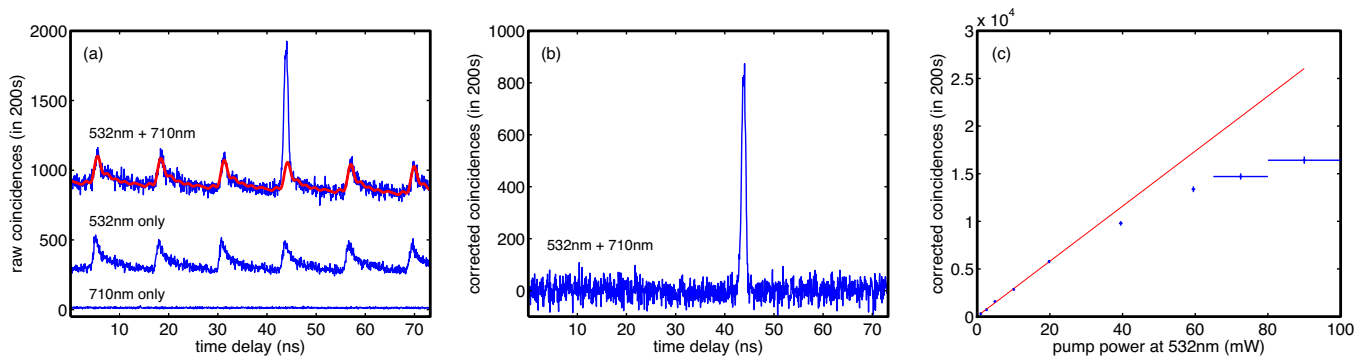}
\end{center}
\caption{Experimental results. (a) Photon doubling signal resulting from the four-mode nonlinear interaction which underlies our four-mode CPC. The signal is only observed when both input beams are present. The strong periodic background is caused by accidental coincidences from single photons created by Raman scattering from both beams, although very few accidentals arise from just the 710 nm input as it creates very few Raman photons at 504 nm. (b) The full MCA trace allows us to correct very precisely for this periodic background and isolate a background-subtracted signal that arises just from the CPC interaction. (c) The pair-production rate depends linearly on the pump power with some saturation at higher pump powers. The saturation arises predominantly from detector and counting saturation and the reduced performance at high powers of the generic single-mode fibres used to spatially filter the 532 nm beam before it is coupled into the PCF. In particular, the detector saturation results mainly from unwanted Raman scattering and it should be possible to dramatically suppress this effect by cooling the PCF.  The linear fit (for the points up to 20 mW of pump) corresponds to a pair detection rate (for 8.5 mW 710 nm) of $1.45 {\pm} 0.02$~pairs per second per mW of 532 nm pump power.}
\label{fig:results}
\end{figure}

The basic experimental design (Fig.~\ref{fig:experiment}) consists of three main stages: 1) state preparation, where the input beams are prepared (polarisation and spatial mode) and combined; 2) nonlinear interaction, where the beams are coupled into the nonlinear PCF for the CPC interaction; and 3) beam analysis, where the beams are separated and analysed after exiting the PCF.  Figure~\ref{fig:results}a illustrates the signal observed for around 90 mW of 532 nm and 8.5 mW of 710 nm average power in the fibre (both horizontally polarised), with both detectors set to analyse V.  The diode laser delivers around $10^5$ photons per pulse during the 532 nm pulses, which are more than three orders of magnitude stronger.  The background signals measured when only one beam is present show clearly that the observed peak is a combined effect of both input beams.  The full time trace allows us to correct very precisely for the periodic background and isolate the signal that arises just from the CPC interaction (Fig.~\ref{fig:results}b).  Polarisation and spectral measurements (see Supplementary Information) show that the interaction is strongly polarisation sensitive, making it suitable for implementing a CZ gate directly in the polarisation degree of freedom, and that birefringent phase-matching ensures that the output photons emerge in two distinct spectral bands.  Together, these measurements confirm that the observed photon pairs result from the target four-colour nonlinear interaction which underlies our four-mode CPC process.

Figure~\ref{fig:results}c shows the dependence of the pair-production rate on the pump power, which exhibits a linear trend with exponential saturation at higher pump powers.  The saturation arises predominantly from two technical effects, detector and counting saturation and reduced performance at high powers of generic single-mode fibres for spatial filtering, both of which can be addressed in future experiments.  We can now use the results from Fig.~\ref{fig:results}c to experimentally estimate the nonlinear interaction strength, $\Gamma t$, that appears in Eq.~\ref{eq:evolution-onephoton}.  The linear fit (for the points up to 20 mW of pump) corresponds to a pair detection rate (for 8.5 mW 710 nm) of $1.45 {\pm} 0.02$~pairs per second per mW of 532 nm pump power.  Taking into account the measured losses due to coupling and optical elements in the beam analysis circuit (${\sim}2.6\%$ in arm 1; ${\sim}14.6\%$ in arm 2), this corresponds to a nonlinear interaction parameter  \emph{inside the PCF} of $\Gamma t \sim 5 \times 10^{-6}$ per $\sqrt{\rm mW}$ (estimated directly from Eq.~\ref{eq:evolution-onephoton}).  For a reasonable fibre-coupled pump power of 1 W, this gives $\Gamma t \sim 10^{-4}$.  We might also expect to improve this by one or two orders of magnitude in future experiments with technical improvements, by specifically engineering the nonlinear and dispersion properties of longer PCFs and matching them with the optical wavelengths.  Therefore, while reaching the deterministic regime ($\Gamma t \sim \pi/2$) using silica might be challenging, it should be feasible with current PCF technology based on other materials such as chalcogenide glasses, where the $\chi^{(3)}$ nonlinearity (which is proportional to $\gamma$) is around $10^3$ times larger than in silica~\cite{EggletonBJ2011a}.


\section{Discussion}

Coherent photon conversion is a classically pumped nonlinear process which produces coherent oscillations between orthogonal states involving multiple quantum excitations, providing a new way to generate and process complex multiquanta states.  Using higher-order nonlinearities with multiple pump fields also allows a mechanism for multiparty mediation of the dynamics.

One special case, based around a single pumped four-wave mixing process, on its own already provides a versatile array of tools which could have significant impact as building blocks for many quantum technologies, including quantum computing.  In particular, we show that two of these building blocks are already sufficient to define a new CPC-based approach for photonic quantum computing that fulfills all of the DiVincenzo criteria, including providing deterministic two-qubit entangling gates based on a novel type of effective photon-photon interaction induced by Berry's phase effects, heralded multiphoton sources with almost no higher-order terms and efficient, low-noise single-photon detection using real-world detectors.
Importantly, CPC could also provide strong benefits for optical QIP experiments even for interaction strengths substantially less than is required for deterministic operation.  For example, the photon doubler provides improved performance for single-photon detection, even when less than 100\% efficient.  Also, the photon doubler does not introduce any of the extra higher-order terms that limit the performance of down-conversion based photon sources~\cite{BarbieriM2009a}.  As a result, even at low efficiencies, a CPC-based multiphoton source offers the potential for higher multiphoton rates with much lower noise terms.

Our experiments provide a proof-of-principle demonstration of the process underlying four-mode CPC, demonstrating that an effective $\chi^{(2)}$ nonlinearity can be produced and tuned in a material where such a nonlinearity in otherwise unavailable.  The results also suggest that operation efficiencies near 100\% could be achieved by using current, albeit sophisticated PCF technology, e.g., based on chalcogenide glasses~\cite{EggletonBJ2011a}.  The availability of such nonlinearities in the optical regime would open the possibility of large-scale QIP with single and entangled photons. Furthermore, since four-wave CPC is derived from a $\chi^{(3)}$ nonlinear interaction where near-degenerate operation is possible, this process is therefore compatible with telecom technology (unlike normal $\chi^{(2)}$ processes) and is also ideally suited to integrated optics and waveguide applications.

Finally, since our scheme is based only on interacting bosonic fields, its applications are not restricted to optical systems.  It could also be implemented in optomechanical, electromechanical and superconducting systems where extremely strong intrinsic nonlinearities involving vibrational~\cite{HolmesCA2009a,ChangDE2011a,ThompsonJD2008a} or matter-based~\cite{MoonK2005a,MarquardtF2007a} degrees of freedom are more readily available.

\section*{Acknowledgements}

The authors would like to acknowledge helpful discussions with T.~Jennewein, A.~Fedrizzi, D.~R.~Austin, T.~Paterek, B.~J.~Smith, W.~J.~Wadsworth, M. Halder, J.~G.~Rarity, F.~Verstraete and A.~G.~White.  This work was supported by the ERC (Advanced Grant QIT4QAD), the Austrian Science Fund (Grant F4007), the EC (QU-ESSENCE and QAP), the Vienna Doctoral Program on Complex Quantum Systems (CoQuS), the John Templeton Foundation and also in part by the Japanese FIRST program.

$^*$ Corresponding authors: NKL (email: nathan.langford@univie.ac.at) or AZ (email: anton.zeilinger@univie.ac.at).

\appendix

\section{Supplementary Information}

\subsection{Scaling up.}

In order to achieve scaleable operation, we need a small number of basic units which can be simply connected together in an appropriately designed network.  For example, the wavelengths of the photons at the outputs of each unit should be compatible with the input modes of the next unit.  Here, we suggest two different approaches to achieve this goal.  Both methods are based entirely on processes that use the same type of four-wave CPC interaction and can in principle be made 100\% efficient.  Finally, both methods can also operate with four near-degenerate frequency modes and are therefore compatible with the extensive standard telecommunications toolbox.

\emph{Method 1} (see Fig.~1b of main text), as explained already in the main text, builds the photon-doubling units from two stages, both of which utilise the same CPC interaction.  The first stage uses the basic CPC interaction to take a single photon at $\omega_a$ and create single photons at $\omega_b$ and $\omega_c$.  The next stage modifies the same interaction simply by adding an extra pump beam at either $\omega_b$ or $\omega_c$, thus creating a process which implements single-photon frequency conversion between the remaining mode ($c$ or $b$) and the original mode, $a$.  Thus, using the same CPC interaction and the same high-power pump at $\omega_d$, with the addition of two relatively low-power pumps at $\omega_b$ and $\omega_c$, the output photons from the photon doubler can be individually converted back into photons at the original frequency, $\omega_a$.  The required resources for this process scale linearly with the number of output photons (one CPC photon doubler and two CPC frequency converters per extra photon).  The controlled-phase gate illustrated in Fig.~1a of the main text, which takes input photons at frequencies $\omega_b$ and $\omega_c$, can be modified in a similar way using frequency conversion to build a unit which implements a controlled-phase gate between input photons at $\omega_a$.  Of course, in ``compiling'' any larger network of such units, many of these frequency conversion stages are redundant and can be removed to minimise the total number of nonlinear interaction steps used.

\begin{figure}
\begin{center}
\includegraphics[width=0.85\columnwidth]{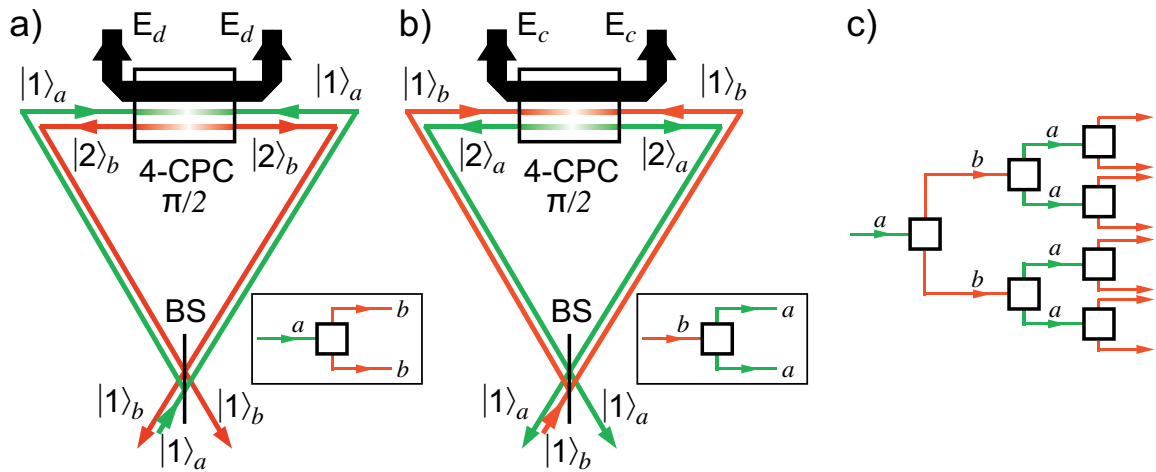}
\end{center}
\caption{Method 2: a) Single photon doubler for mode $a$, producing two photons in the degenerate mode $b$. Separating the two photons into different modes is achieved via a ``reverse Hong-Ou-Mandel''-type interaction at a beam splitter ($|2,0\rangle + |0,2\rangle \rightarrow |1,1\rangle$). With the pump field, $E_c$, energy conservation is given by $\omega_a + \omega_c = 2 \omega_b$. b) Analogous photon doubler for mode $b$, but with the roles of modes $a$ and $b$ swapped ($a$ now degenerate). With the different pump field, $E_d$, energy conservation is now given by $\omega_b + \omega_d = 2 \omega_a$. c) Cascaded concatenation of a) and b) to achieve scaling up to high number of photons.}
\label{fig:scalingup-method2}
\end{figure}

\emph{Method 2} takes a slightly different approach (Fig.~\ref{fig:scalingup-method2}).  So far, we have considered only the CPC process involving four distinct frequency modes (described by the Hamiltonian $H = \gamma a^\dag bc d^\dag + \gamma^\ast a b^\dag c^\dag d$), but it is also possible to implement a special case of this process when the two modes $b$ and $c$ are degenerate.  Indeed, this is just the standard four-wave mixing interaction used in most spontaneous four-wave mixing (SFWM) sources.  In this case, the full Hamiltonian is:
\begin{equation}
H = \gamma a b^{\dag 2} d + \gamma^\ast a^\dag b^2 d^\dag,
\end{equation}
and when mode $d$ is pumped by a bright classical laser beam, $E_d$, the effective Hamiltonian reduces to:
\begin{equation}
\tilde{H} = \tilde{\gamma} a b^{\dag 2} + \tilde{\gamma}^\ast a^\dag b^2.
\end{equation}
This modified, degenerate form of CPC also implements a photon doubling process, but takes a single-photon input at $\omega_a$ and produces instead a two-photon Fock-state in $\omega_b$.  Then, by embedding this in two arms of an interferometer, the two degenerate and indistinguishable photons can be deterministically separated into different modes via a ``reverse Hong-Ou-Mandel''-type interaction at a beam splitter ($|2,0\rangle + |0,2\rangle \rightarrow |1,1\rangle$).  Each of these photons (at $\omega_b$) can then in turn be converted in a similar way into two photons at $\omega_a$ using the same CPC interaction with a different high-power pump frequency.  By alternating between these two processes, the photon doubling cascade can therefore be scaled up to larger systems.  As with Method 1, this approach is built using units with the same type of CPC interaction, this time simply with two different pump frequencies.  Once again, this scaled process requires only linear resources to create $n$-photon states (3 CPC photon doublers per extra 3 photons).

\subsection{Entanglement sources.}

\begin{figure}
\begin{center}
\includegraphics[width=0.6\columnwidth]{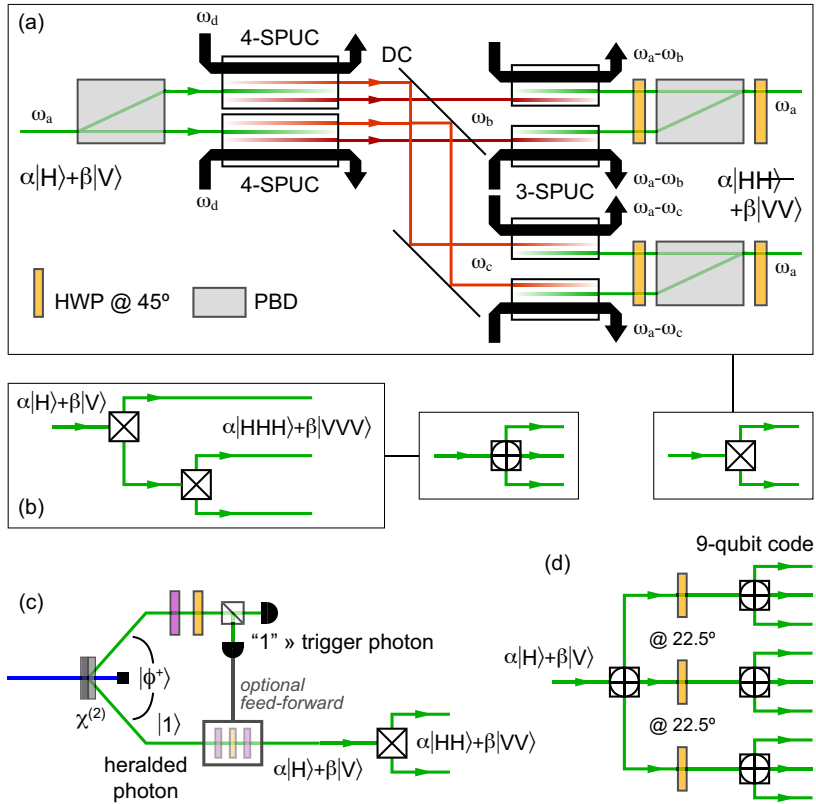}
\end{center}
\caption{CPC circuits for creating (a) bipartite and (b) tripartite entanglement.  (c) A CPC circuit for generating arbitrary, nonlocally prepared, nonmaximally entangled states.  (d) A direct implementation of the error-correction encoding step for a simple 9-qubit code.}
\label{fig:entanglementsources}
\end{figure}

The cascaded photon-doubling technique for efficiently generating multiphoton states is not limited to generating simple pure product states.  Figures~\ref{fig:entanglementsources}a and~\ref{fig:entanglementsources}b illustrate CPC circuits for generating both bipartite and genuine tripartite entanglement.  For example, if two polarisation states of an input single photon are split up in a polarisation-sensitive interferometer (e.g., a Mach-Zender interferometer using polarising beam displacers, or a Sagnac source~\cite{FedrizziA2007a}) and if a CPC photon doubling interaction is applied to both polarisations, then the same process can be used as a source of polarisation-entangled photons (Fig.~\ref{fig:entanglementsources}a).  The same result could also be achieved using a ``crystal-sandwich'' arrangement.  Existing experiments have already repeatedly demonstrated that these techniques produce extremely high-quality entanglement, and by tuning the polarisation of the input single photon, we can generate different (both maximally and non-maximally) entangled states.  This state can even be prepared nonlocally (Fig.~\ref{fig:entanglementsources}c).  Finally, Fig.~\ref{fig:entanglementsources}d illustrates how these basic building blocks can be combined to directly implement the encoding step of an error-correction protocol using a simple 9-qubit code.

\subsection{Efficient detectors.}

\begin{figure}
\begin{center}
\includegraphics[width=0.8\columnwidth]{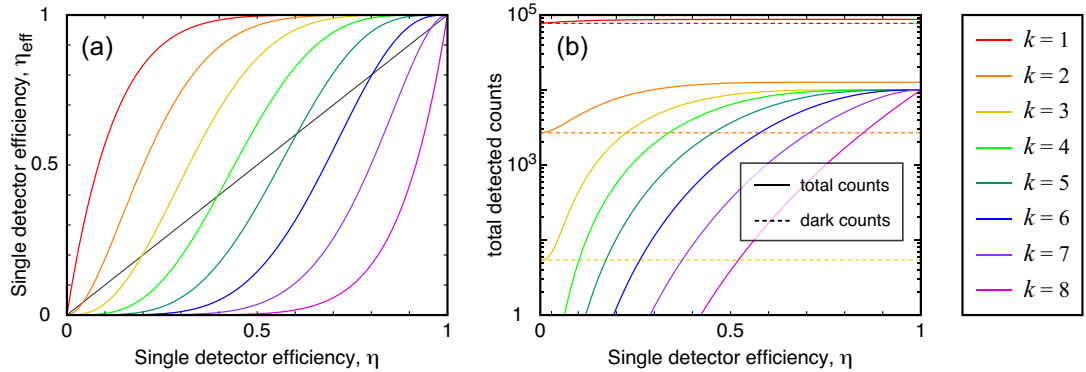}
\end{center}
\caption{Efficient detection via a photon-doubling cascade. (a) Effective efficiency, $\eta_{\rm eff}$, for detector cascade with 3 cascade steps ($n{=}8$) and a $k$-fold coincidence ``clicks'' between $k{=}1$ (red) and $k{=}8$ (violet).  (b) Predicted counts for a simulated probabilistic experiment with $10^6$ rep.\ rate, $10^{-2}$ incident photon probability and $10^{-2}$ (individual detector) dark count probability (i.e., SNR=1), using a detector cascade with $n{=}8$ and $k$ between 1 (red) and 8 (violet).}
\label{fig:efficient-detection}
\end{figure}

Figure~\ref{fig:efficient-detection}a shows the effective detector efficiency for a 3-step cascade, which produces 8 ($2^3$) photons, and various $k$, where $k$-fold coincidences are defined to signal successful detection events. When $k=1$, it is clear that the detector efficiency can be greatly increased, although there is naturally some trade-off when coincidence detection is used to suppress the dark counts. Figure~\ref{fig:efficient-detection}b illustrates how this scheme would improve both count rates and signal-to-noise ratio in an example where the single-detector dark counts are as large as the signal.  Interestingly, without using coincidence detection to suppress it, the effective dark count noise is actually higher for the cascade than for an individual detector. This effect is quite pronounced in the example shown, because we deliberately chose an extremely high dark count rate, but it is already overcome with $k{=}2$ only.  In many practical situations, however, the raw dark count probability is much less than the signal rate and this effect is not significant.

\subsection{Higher-order applications: Fock-state preparation.}

\begin{table}
\renewcommand{\arraystretch}{0.65}
\begin{center}\begin{tabular}{ccc}
\hline\hline
target & interaction & target \\
state & $\Gamma\tau$ ($\times \pi$) & transmission \\ \hline\hline
$|1\rangle$ & 1 & 1.000 \\ \hline
$|2\rangle$ & $2/\sqrt{6}$ & 1.000 \\ \hline
$|3\rangle$ & 5.805 & $>0.9998$ \\ \hline
$|4\rangle$ & 2.154 & $>0.9999$ \\ \hline
$|5\rangle$ & 21.278 & $>0.990$ \\ \hline
$|6\rangle$ & 11.100 & $>0.996$ \\ \hline
$|7\rangle$ & 9.390 & $>0.986$ \\
& 68.972 & $>0.995$ \\ \hline
$|8\rangle$ & 20.024 & $>0.93$\\ \hline
\hline
\end{tabular}\end{center}
\sffamily\caption{Revival peaks for higher-order input states (see text for details).}
\label{tab:fock-filtration}
\end{table}

Table~\ref{tab:fock-filtration} shows interaction lengths and target-state transmission probabilities for some early ``revival'' peaks from different photon-number input states, which have the potential to be used for Fock-state filtration.  Because of the complexity of higher-order eigenvalues, longer interactions are generally required before a significant revival occurs, which is even then generally not 100\% efficient.  This lowers the success probability of the Fock-state preparation for higher-order states, but for lower orders, quite pure Fock states can be prepared non-deterministically with relatively few interaction steps and quite high probability.  For example, as mentioned in the main text, with ${\sim}56\%$ probability, a single-photon state can be prepared with fidelity and purity greater than $99.6\%$ using a weak coherent input state ($|\alpha|^2=1.5$) with only 5 steps.

\subsection{Higher-order applications: improved down-conversion.}

\begin{figure}
\begin{center}
\begin{tabular}{ccc}
\includegraphics[width=0.85\columnwidth]{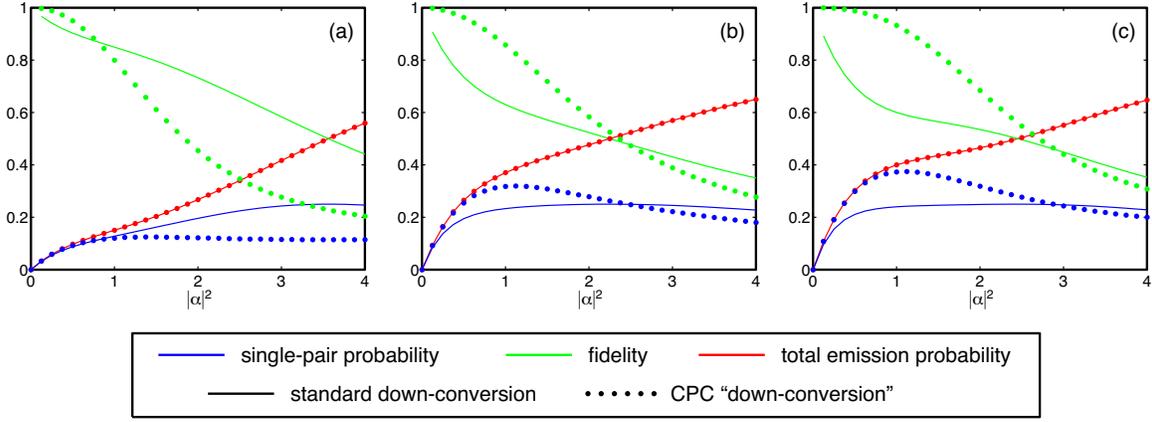}
\end{tabular}
\end{center}
\caption{Improved down-conversion using the CPC interaction.  Overall single-pair probability per pulse (blue) and fidelity of a ``heralded'' state with the one-pair state (green) for a particular total photon-emission probability per pulse (red). Solid circles show the results for interaction times: (a) $\Gamma\tau{=}2\pi/\sqrt{6}$, (b) $\Gamma\tau{=}4\pi/\sqrt{6}$ and (c) $\Gamma\tau{=}6\pi/\sqrt{6}$.  For comparison, the lines show the results for standard down-conversion with the same herald detection probability.}
\label{fig:betterDC}
\end{figure}

We now briefly outline how the CPC interaction can be used to implement an improved form of down-conversion.  Specifically, consider an input state to mode $a$ that contains higher-order Fock states, such as a weak coherent state or the heralded single-photon state created by triggering from a down-conversion pair, and assume that the interaction time corresponds to an integral number of complete oscillations of the $|200\rangle$ input state (i.e., $\Gamma\tau{=}2m\pi/\sqrt{6}$).  In such a situation, the $|100\rangle$ term will have converted to $|011\rangle$ with some finite probability, but the $|200\rangle$ will have remained unconverted with $100\%$ probability, allowing the creation of correlated ``down-conversion-like'' photon pairs, with no $|022\rangle$ term.  Figure~\ref{fig:betterDC} shows the output of this process after each of the first three $|200\rangle$ oscillation periods, given weak coherent input states with a range of average photon numbers.  For both two and three periods (Fig.~\ref{fig:betterDC}), the CPC process produces substantially higher pair-emission probabilities with much higher fidelity than a standard down-conversion source with comparable emission rates.  This technique can also provide improved higher-order characteristics using a heralded single photon from standard down-conversion as the input.

\subsection{Polarisation and spectral dependence of the CPC interaction.}

\begin{figure}
\begin{center}
\includegraphics[width=0.27\columnwidth]{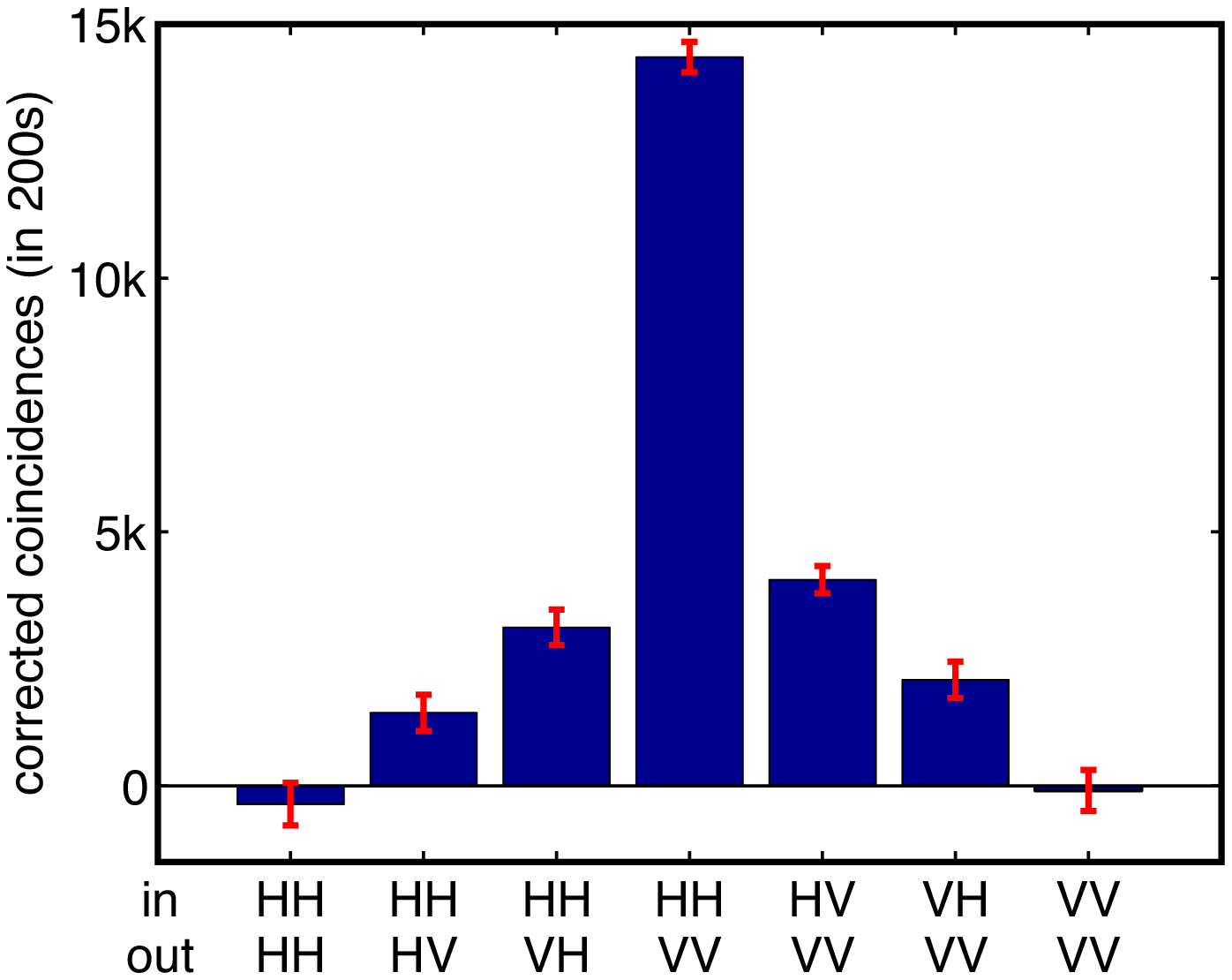}
\includegraphics[width=0.27\columnwidth]{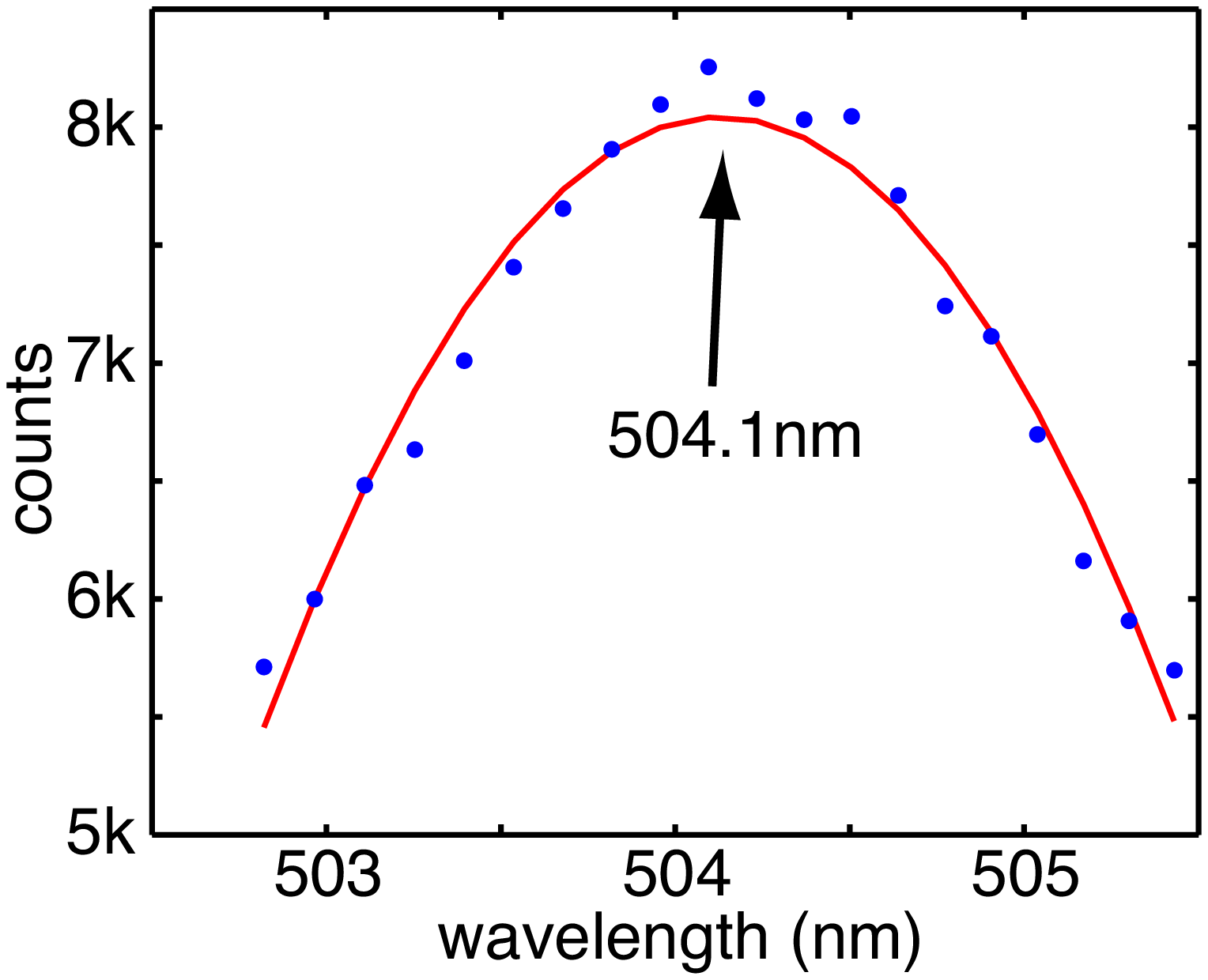}
\includegraphics[width=0.27\columnwidth]{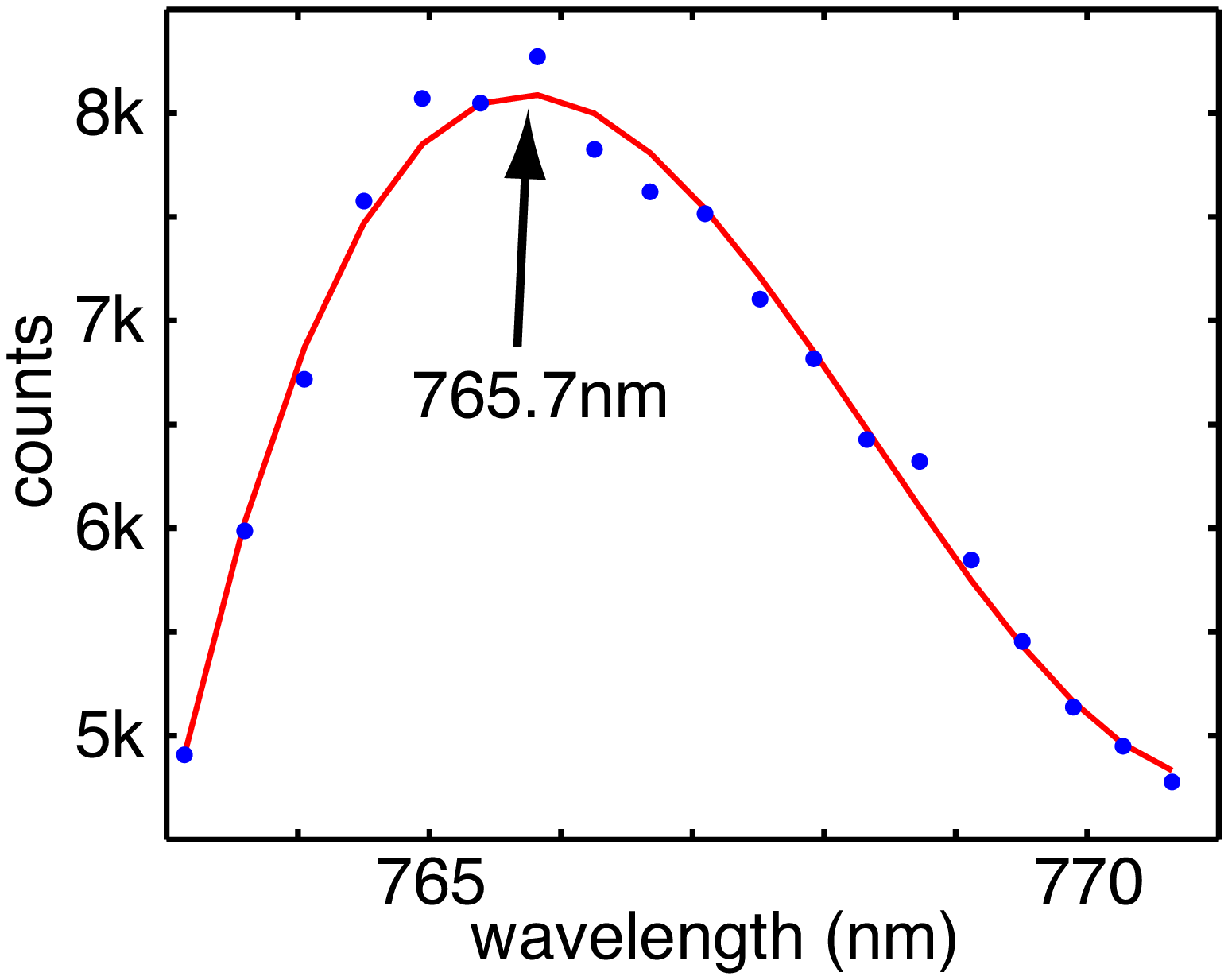}
\end{center}
\caption{(a) Polarisation and (b,c) spectral dependence of the CPC interaction.}
\label{fig:results-poln-spectral}
\end{figure}

Figure~\ref{fig:results-poln-spectral}a shows the polarisation dependence of our four-wave mixing CPC interaction.  The phase-matching in the PCF should ensure that the process can only take place for the correct combination of polarisations.  The results show that the target polarisation clearly gives the strongest signal, whereas combinations with equal polarizations (HHHH and VVVV) for all modes yield no coincidences within statistical error.  For the other measured combinations, however, the signal did not completely vanish. We believe that this arises from a combination of two effects, namely imperfect orientation of the polarisation-maintaining fibre FC-PC connectors and unwanted polarisation cross-talk for the modes below the single-mode cut-off wavelength (${\sim} 650nm$).  In other words, the signals still result from the target CPC interaction, but shows up in the incorrect polarisation because of polarisation cross-talk in the fibre either before or after the interaction.

Figures~\ref{fig:results-poln-spectral}b and c show the spectral and phase-matching dependence of our CPC interaction.  Our single-photon spectrometers were constructed by placing rotating narrow bandpass interference filters (two per spectrometer) in front of multimode fibres.  They were calibrated against a known spectrometer using bright light.  The spectra were integrated for 20 seconds per point using 60 mW in the 532 nm beam.  The coincidence detection for these measurements was performed using a single-channel analyser, with the result that the detected rates could not be corrected for background.  The resulting central wavelengths agree well with the predicted values from our phase-matching calculations.

\bibliography{abbreviations,paper}

\end{document}